\documentstyle[12pt,epsfig]{article}
\def\beqa{\begin{eqnarray}}
\def\eeqa{\end{eqnarray}}
\def\beq{\begin{equation}}
\def\eeq{\end{equation}}

\def\pl{{\it Phys. Lett.}\ }

\def\apj{{\it Ap. J.}\ }

\def\ncb{{\it Il Nuovo Cimento ``B''}}

\def\araa{{\it Ann. Rev. Astr. Ap.}\ }

\def\ie{{\it i.e. }}
\def\eg{{\it e.g. }}

\begin{document}

\begin{titlepage}
  \title{Probing the nature of  compact dark object at the Galactic Center
by gravitational lensing}

\author{S. Capozziello$^{1}$\thanks{
E-mail:capozziello@vaxsa.csied.unisa.it} and G.
Iovane$^{1}$\thanks{E-mail:geriov@vaxsa.csied.unisa.it}\\ {\em
$^{1}$Dipartimento di Scienze Fisiche "E.R. Caianiello",}\\ {\em
INFN Sez. di Napoli, Gruppo collegato di Salerno}\\ {\em
Universit\`a di Salerno, I-84081 Baronissi (SA) Italy.}}
\date{}
\maketitle

\begin{abstract}
The long--standing issues of determination of the mass
distribution and nature of the center of our Galaxy could be
probed by a lensing experiment capable of testing the spatial and
velocity distributions of stars nearby and beyond it. We propose a
lensing toy--model which could be a further evidence that  a
massive consensation (\eg a neutrino condensation) is a good
candidate to explain the data ruling out the presence of a
supermassive black hole.
\end{abstract}

\vspace{20. mm} Keywords:
 supermassive neutrino star, dark matter, gravitational lensing.

          \vfill
          \end{titlepage}

The puzzle to explain the nature of matter condensation at the
center of our Galaxy is more than twenty years old problem
\cite{oort}. Various observational campaigns \cite{genzel1} have
identified such a center with the supermassive compact dark object
Sagittarius A$^*$ (Sgr A$^*$) which is an extremely loud radio
source. Detailed information comes from dynamics of stars moving
in the gravitational field of such a central object. The
statistical properties of spatial and kinematical distributions
are of particular interest \cite{sellgreen}: Using them, it is
possible to establish the mass and the size of the object which
are $(2.61\pm 0.76)\times 10^6 M_{\odot}$ concentrated within a
radius of 0.016 $pc$ (about 30 $lds$)\cite{ghez},\cite{genzel96}.

More precisely, Ghez {\it et al.} \cite{ghez} have made a campaign
of observations where velocity measurements in the central
arcsec$^2$ are extremely accurate.  From this bulk of data, it is
possible to state that a supermassive compact dark object is
present at the center of Galaxy and, furthermore, it is revealed
by the motion of stars moving within a projected distance of less
than 0.01 $pc$ from the radio source Sgr A$^*$ at projected
velocities in excess of 1000 $km/s$.  In other words, a high
increase of velocity dispersion of the stars toward the dynamic
center is revealed. Furthermore, a large and coherent
counter--rotation, expecially of the early--type stars, is
revealed, supporting their origin in a well--defined epoch of star
formation. Besides, observations of stellar winds nearby Sgr A$^*$
give a mass accretion rate of ${\displaystyle
\frac{dM}{dt}=6\times 10^{-6}M_{\odot}yr^{-1}}$ \cite{genzel96}.
Hence, the dark mass must have a density $\sim 10^9
M_{\odot}pc^{-3}$ or greater and a mass--to--luminosity ratio of
at least $100M_{\odot}/L_{\odot}$. The conclusion is that the
central dark mass is statistically very significalt $(\sim
6-8\sigma)$ and cannot be removed even if a highly anisotropic
stellar velocity dispersion is assumed. Given that the majority of
stars in a cluster are of solar mass, such a large density
contrast excludes that the dark mass could be a cluster of almost
$2\times 10^6$ neutron stars or white dwarfs. As a first
conclusion, several authors state that in the Galactic center
there is either a single supermassive black hole or a very compact
cluster of stellar size black holes \cite{genzel96}. The first
hypothesis is supported by several authors since similar
supermassive black holes have been inferred to explain the central
dynamics of several galaxies as M87 \cite{ford},\cite{macchetto},
or NGC4258 \cite{greenhill}.  However, due to the above mentioned
mass accretion rate, if Sgr A$^*$ is a supermassive black hole,
its luminosity should be more than $10^{40}erg\,s^{-1}$. On the
contrary, observations give a bolometric luminosity of
$10^{37}erg\,s^{-1}$. This discrepancy is the so--called
``blackness problem'' which has led to the notion of a ``black
hole on starvation'' at the center of Galaxy. Besides, the most
recent observations probe the gravitational potential at a radius
larger than $4\times 10^{4}$ Schwarzschild radii of a black hole
of mass $2.6\times 10^{6}M_{\odot}$ \cite{ghez} so that the
supermassive black hole hypothesis at the center of Galaxy is far
from being conclusive.

On the other hand, stability criteria rule out the hypothesis of a
very compact stellar cluster in Sgr A$^{*}$ \cite{sanders}. In
fact, detailed calculations of evaporation and colision mechanisms
give maximal lifetimes of the order of $10^8$ years which are much
shorter than the estimed age of the Galaxy \cite{maoz}.

Another viable and, in some sense more attractive alternative
model for the supermassive compact object in the center of our
Galaxy (and in the center of several other galaxies) has been
recently proposed by Viollier {\it et al.} \cite{viollier}. The
main ingredient of the proposal is that the dark matter at the
center of galaxies is made by nonbaryonic matter (\eg massive
neutrinos or gravitinos) which interacts gravitationally forming
supermassive balls in which the degeneracy pressure of fermions
balances their self--gravity. Such neutrino balls could have
formed in the early epochs during a first--order gravitational
phase transition and their dynamics could be reconciled with some
adjustments to the Standard Model of Cosmology (for an exhaustive
discussion of the problem, see \cite{viollier}).

Furthermore, several experiments are today running to search for
neutrino oscillations. LSND \cite{lsnd} finds evidence for
oscillations in the $\nu_{e}-\nu_{\mu} $ channel for pion decay at
rest and in flight. On the contrary KARMEN \cite{karmen} seems to
be in contradiction with LSND evidence. CHORUS and NOMAD at CERN
are just finished the phase  of ('94--'95) data analysis. In any
case, it is very likely that exact preditions for and
$\nu_{\mu}-\nu_{\tau}$ oscillations will be available at the end
of millennium or in first years of the next.

From all this bulk of data, and thanks to the fact that it is
possible to give correct values for the masses to the quarks {\it
up}, {\it charm}, and {\it top}, it is possible to infer
reasonable values of mass for $\nu_{e}$, $\nu_{\mu}$, and
$\nu_{\tau}$. For our purposes, we are particularly interested in
fermions which masses range between 10 and 25 keV$/c^{2}$
which cosmologically fall into the category of {\it warm} dark matter%
\footnote{ A good estimated value for the mass of $\tau$-neutrino
is
\[
m_{{\nu_{\tau}}}=m_{{\nu_{\mu}}}\left(\frac{m_{t}}{m_{c}}\right)^{2}
\simeq 14.4 \mbox{keV}/c^{2}\,,
\]
which well falls into the above range.}. Choosing fermions like
neutrinos or gravitinos in this mass range allows the formation of
supermassive degenerate objects (from $10^6 M_{\odot}$ to $10^9
M_{\odot}$). As we said, the existence of such objects avoids to
invoke the supermassive black hole hypothesis in the center of
galaxies and quasars and it is able to justify the large amount of
radio emission coming from such unseen objects.

The theory of heavy neutrino condensates, bound by gravity, can be
easily sketched \cite{leimgruber}. Let us consider the
Thomas--Fermi model for fermions. We can set the Fermi energy
$E_{F}$ equal to the gravitational potential which binds the
system, that is
\begin{equation}  \label{n1}
\frac{\hbar^2 k_{F}^{2}(r)}{2 m_{\nu}}-m_{\nu}\Phi(r)=E_{F}
=-m_{\nu}\Phi(r_{0})\,,
\end{equation}
where $\Phi(r)$ is the gravitational potential, $k_{F}$ is the
Fermi wave number and $\Phi(r_0)$ is a constant chosen to cancel
the gravitational potential for vanishing neutrino density. The
length $r_{0}$ is the estimed size of the halo. If we take into
account a degenerate Fermi gas, we get
$k_{F}(r)=\left(6\pi^{2}n_{\nu}(r)/g_{\nu}\right)^{1/3},$ where
$n_{\nu}(r)$ is the neutrino number density and we are assuming
that it is the same for neutrinos and antineutrinos within the
halo. The number $g_{\nu}$ is the spin degeneracy factor.
Immediately we see that the number density is a function of the
gravitational potential, {\it i.e. } $n_{\nu}=f(\Phi),$ and the
model is specified by it. If in the center of the neutrino
condensate there is a baryonic star (which we approximate as a
point source), the gravitational potential will obey a Poisson
equation where neutrinos (and antineutrinos) are the source term,
{\it i.e. }
\begin{equation}  \label{n4}
\triangle\Phi=-4\pi Gm_{\nu}n_{\nu}\,.
\end{equation}
Such an equation is valid everywhere except at the origin. We can
assume,
for the sake of simplicity, the spherical symmetry and define the variable $%
u=r[\Phi(r)-\Phi(r_{0})]$ then the Poisson equation reduces to the
radial Lan\'e--Emden differential equation
\begin{equation}  \label{n5}
\frac{d^2 u}{dr^2}= -\left(\frac{4\sqrt{2}m_{\nu}^{4}G g_{\nu}}{3\pi\hbar^{3}%
}\right) \frac{u^{3/2}}{\sqrt{r}}\,,
\end{equation}
with polytropic index $n=3/2$. This equation is equivalent to the
Thomas--Fermi differential equation of atomic physics, except for
the minus sign that is due to the gravitational attraction of the
neutrinos as opposed to the electrostatic repulsion between the
electrons. If $M_{B}$ is the mass of the baryonic star internal to
the condensation, the natural boundary conditions are
\begin{equation}
u(0)=GM_{B}\,,\;\;\;\;\;\;\;u(r_{0})=0\,.
\end{equation}
Recasting the problem in a dimensionless form, we have
\begin{equation}  \label{n6}
\frac{d^{2}v}{dx^2}=-\frac{v^{3/2}}{\sqrt{x}}\,,
\end{equation}
and the boundary conditions
\begin{equation}  \label{n7}
v(0)=\frac{M_{B}}{M_{\odot}}\,,\;\;\;\;\;\;\;v(x_{0})=0\,,
\end{equation}
with the positions
\begin{equation}  \label{n8}
v=\frac{u}{GM_{\odot}}\,,\;\;\;\;\;\;\;\;x=\frac{r}{a}\,,
\end{equation}
and
\begin{equation}  \label{n9}
a=\left(\frac{3\pi \hbar^3}{4\sqrt{2}m_{\nu}^4g_{\nu} G^{3/2}M_{\odot}^{1/2}}%
\right)^{2/3}= 2.1377\left(\frac{17.2\mbox{keV}}{m_{\nu}c^2}%
\right)^{8/3}g_{\nu}^{-2/3} \mbox{lyr}\,.
\end{equation}
We have to note that for $m_{\nu}=17.2\mbox{keV}/c^2$ the
characteristic scale $a$, the corresponding of the Bohr radius for
an electron bound to a nucleus, is, for a neutrino halo bound by a
baryonic star, of the order of the average distance between stars.
It strongly depends on neutrino mass.

All the quantities characterizing the condensate can be written in
terms of $v$ and $x$ (or $u$ and $r$):
\begin{equation}  \label{n10}
\Phi(r)=\Phi(r_{0})+\frac{u}{r}\,,\;\;
n_{\nu}(r)=\frac{m_{\nu}^3 g_{\nu}}{6\pi^2\hbar^3} \left(\frac{2u}{r}%
\right)^{3/2}\,,\;\;
P_{\nu}(r)=\left(\frac{6}{g_{\nu}}\right)^{2/3} \frac{\pi^{4/3}\hbar^{2}}{%
5m_{\nu}}n_{\nu}(r)^{5/3}\,,
\end{equation}
which are, respectively, the gravitational potential, the number
density, and the degeneracy pressure. As shown in \cite{viollier},
the general solution of (\ref{n5}), or equivalently (\ref{n6}),
has scaling properties and it is able to reproduce the
observations. In particular, it could well fit the observations
toward the center of our Galaxy which estimate, considering the
proper motion ($\leq 20$ km sec$^{-1}$) of the source Sgr A$^*$, a
massive object of $M=(2.6\pm 0.7)\times 10^{6}M_{\odot}$ which
dominates the gravitational potential in the inner ($\leq 0.5$pc)
region of the bulge \cite{sagitta}. In summary a degenerate
neutrino star
of mass $M=2.6\times 10^{6}M_{\odot}$, consisting of neutrino with mass $%
m\geq 12.0$ keV$/c^{2}$ for $g_{\nu}=4$, or $m\geq 14.3$ keV$/c^{2}$ for $%
g_{\nu}=2$, does not contradict the observations. Considering a
standard accretion disk, the data are in agreement with the model
if Sgr A$^*$ is a neutrino star with radius $R=30.3$ ld ($\sim
10^5$ Schwarzschild radii) and
mass $M=2.6\times 10^{6}M_{\odot}$ with a luminosity $L\sim 10^{37}$erg sec$%
^{-1}$.

Similar results hold also for the dark object ($M\sim 3\times
10^{9}M_{\odot} $) inside the center of M87.

Now the problem is: How much is the model consistent? Could it be
improved at boundaries? Actually, due to the Thomas--Fermi theory,
the model fails at the origin and, in any case, we have to
consider the effect of the surrounding baryonic matter which, in
some sense, have to give stability to
the neutrino condensate. In fact, an exact solution of Eq.(\ref{n5}) or Eq.(%
\ref{n6}) is $u(r)\sim r^{-3}$ from which $\Phi(r)\sim r^{-4}$
which is clearly unbounded from below.

Let us now assume a thermodynamical phase where a constant
neutrino number density can be taken into consideration. This is
quite natural for a Fermi gas at temperature $T=0$. The Poisson
equation is
\begin{equation}  \label{n13}
\triangle\Phi=-4\pi Gm_{\nu}n_{\nu}=\mbox{cost}\,.
\end{equation}
which, for spherical symmetry, can be recast in a Lan\'e--Emden
form
\begin{equation}  \label{n14}
\frac{1}{r^2}\frac{d}{dr}\left(r^2\frac{d\Phi}{dr}\right)=\mbox{cost}\,,
\end{equation}
with polytropic index $n=0$. The solution of such an equation is
$\Phi(r)\sim r^{2},$ which is clearly bounded. For $T\neq 0$ the
above theory holds so that we get a solution of the form
$\Phi(r)\sim r^{-4}$. Matching the two results it is possible to
confine the neutrino condensate. On the other hand, a similar
result is recovered using the Newton Theorem for a spherically
symmetric distribution of matter of radius $R$ \cite{binney}. In
that case, the potential goes quadratically inside the sphere
while it goes as $\Phi(r)\sim r^{-1}$ matching on the boundary. In
our case, the situation is similar assuming the matching with a
steeper potential.

Assuming the existence of such a neutrino condensate in the center
of Galaxy, it could act as a spherical lens for the stars behind
so that their apparent velocities will be larger than in reality.
Comparing this effects with the proper motion of the stars of the
cluster near Sgr A$^*$, exact determinations of the physical
parameters of the neutrino ball could be possible. In this case,
gravitational lensing, always used to investigate baryonic
objects, could result useful in order to detect a nonbaryonic
compact object. Furthermore, since the astrophysical features of
the object in Sgr A$^*$ are quite well known \cite{genzel96},
accurate observations by lensing could contribute to the exact
determination of particle constituents which could be, for
example, neutrinos or gravitinos. Besides, microlensing by cold
dark matter particles and noncompact objects has been widely
considered in literature \cite{gurevich}, being gravitational
lensing independent of the nature and the physical state of
deflecting mass. In fact, any gravitationally condensed structure
can act, in principle, as a gravitational lens. Our heavy neutrino
ball, being massive, extended and transparent, can be actually
considered as a magnifying glass for stars moving behind it. If an
observer is on Earth and he is looking at the center of our Galaxy
(which is at a distance of 8.5 $Kpc$), he should appreciate a
difference in the motion of stars since lensed stars and
non-lensed stars should have different projected velocity
distributions. In other words, depending on the line of sight
(toward the ball or outside the ball) it should be possible to
correct or not the projected velocities by a gravitational lensing
contribution and try to explain the bimodal distribution actually
observed \cite{ghez},\cite{genzel96}.

Let us discuss the physical reasons why a heavy neutrino ball can
be treated as a thick lens.

For a static gravitational field, the refraction index is
connected to the Newtonian potential by the equation
\begin{equation}
\label{pot} n({\bf r})=1-2\frac{\Phi ({\bf r})}{c^{2}}\,
\end{equation}
easily derived by
\begin{equation}
g_{00}\simeq 1+2\frac{\Phi ({\bf
r})}{c^{2}}\;;\;\;\;\;\;g_{ik}\simeq -\delta _{ik}\left(
1-2\frac{\Phi ({\bf r})}{c^{2}}\right) \;;
\end{equation}
assuming the weak field $\Phi /c^{2}<<1$ and the slow motion approximation $%
\left| v\right| <<c$ \cite{straumann}. In this situation, almost
all the usual geometric optics works. Our neutrino ball gives rise
to a static gravitational field, it is an extended object and, due
to neutrinos, it can be reasonably approximated by a
``transparent'' medium. In this case, it is essentially a thick
spherical lens which can be replaced by two thin lenses at a
distance ``$d$'' \cite{fleury}. It is easy to show, by elementary
optics arguments, that this double dioptric system can be
described by the equation
\begin{equation}
\frac{1}{f}=\frac{1}{f_{1}}+\frac{1}{f_{2}}-\frac{d}{f_{1}f_{2}}\,,
\end{equation}
where $f_{1,2}$ are focal lengths. The relations with the
gravitational field and the size of the neutrino ball are given by
\begin{equation}
\frac{1}{f_{i}}=(n-1)\frac{1}{r_{i}}\,,\;\;\;\;\;\;\;\;\;i=1,2
\end{equation}
and $r_{i}=R\simeq d/2$, where $n\left( r\right) $ is the
refraction index induced by the Newtonian potential $R$ is the
neutrino star radius. Since we are assuming a spherically
symmetric distribution of matter inside a radius $R$, the Newton
theorem holds \cite{binney} so that
\begin{equation}
\Phi(r)=\frac{1}{2}\omega ^{2}r^{2}\,,
\end{equation}
with $\omega ^{2}=4\pi G\rho /c^{2}$. The focal length is then
given by \cite{waveguide}
\begin{equation}
L_{foc}=\sqrt{\frac{\pi c^{2}}{4G\rho }}\,,
\end{equation}
which has the same order of magnitude of $f$ and $f_{1,2}$. If
$M=2.6\times 10^6M_{\odot }$ and $R\simeq 30.3\, ld$,
$L_{f_{oc}}\simeq 70ly$ and we have to expect lensing effects
approximately in this range for the stars behind the ball.
Depending on the radius, Eq.(\ref{pot}) gives the refraction index
so that the model is completely determined. Further considerations
give also the range of validity of paraxial approximation. In
fact, given a ray of light entering in the neutrino ball, it is
possible to calculate, the angle of incidence $\alpha $ of the ray
on the back surface, the angle of deflection $\delta $ and the
entering angle $\eta $, (with respect to the normal) which
produces the minimal deflection. See the Fig.1.

\begin{figure}[h!]
\begin{center}
\epsfig{file=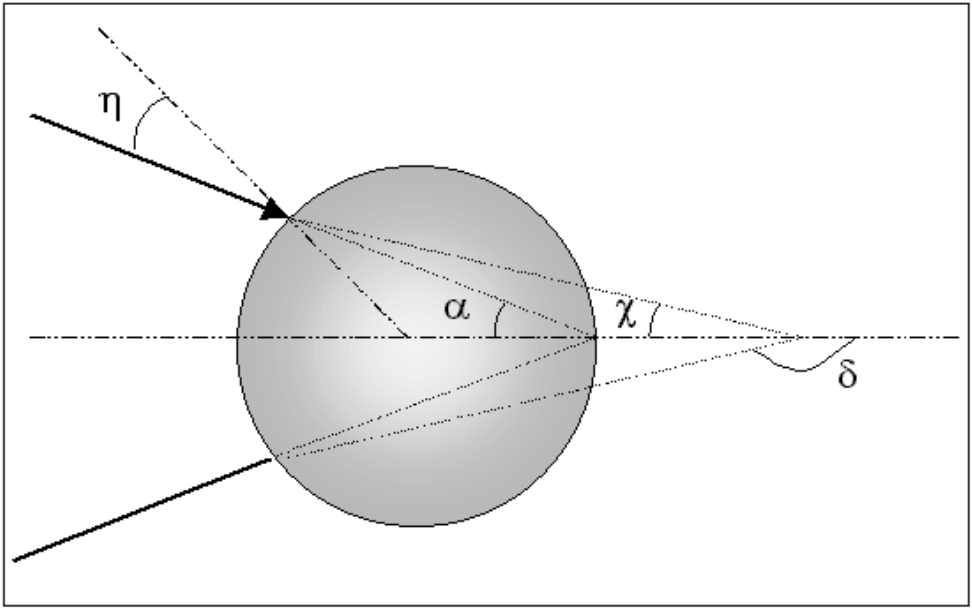,width=13cm,clip=}
\end{center}
\caption{sketch of the lensing effect by a neutrinos ball.}
\label{fig1}
\end{figure}

By knowing these parameters, we could say if an intervening star
undergoes this magnifying glass effect. Snell's law gives $n\sin
\alpha =\sin \eta$ so that the relation
\begin{equation}
\sin \alpha =\frac{1}{n}\sin \eta <\frac{1}{n}\,,
\end{equation}
holds. The angle of incidence $\alpha $ has to be smaller than the
critical angle ${\displaystyle \arcsin \frac{1}{n}}$ so that the
incident light is reflected partially at the back spherical
surface. The net effect is that we should lose a part of the
luminosity of the star population behind the ball. As $\alpha
=(\eta -\alpha )+\chi$, we have
\begin{equation}
\delta =\pi -2\chi =\pi -4\alpha +2\eta ;
\end{equation}
which is the deflection angle. The minimal deflection is deduced
in a straightforward way. We require
\begin{equation}
\frac{d\delta }{d\eta }=-4\frac{d\alpha }{d\eta }+2=0,
\end{equation}
which is ${\displaystyle \frac{d\alpha }{d\eta }=\frac{1}{2}}$.
Being ${\displaystyle \alpha =\arcsin (\frac{1}{n}\sin \eta)}$,
immediately we get ${\displaystyle \frac{d\alpha }{d\eta }
=\frac{\cos \eta }{n\cos \alpha }}$ and then $$ 1-\frac{1}{n}\sin
^{2}\eta =\frac{4}{n^{2}}\cos ^{2}\eta\,. $$ Finally
\begin{equation}
\label{glass} \cos ^{2}\eta =\frac{n^{2}-1}{3},
\end{equation}
which relates the entering incidence angle with the refraction
index. Summing up all these information, we should say if a given
star undergoes or not a significant lensing effect behind the
neutrino ball.

Let us now take into account the projected positions $(x,y)$ in
the sky of the observed early- and late-type stars as reported in
\cite{genzel96}. The gravitational refraction index as given by
Eq.(\ref{pot}) for a  supermassive neutrino of mass $M=2.6\times
10^{6}M_{\odot }$ and radius $R=30.3\,ld$ is $n=1+5\times
10^{-6}$. From the above considerations, the angles $\eta $ and
$\alpha $ are given by
\begin{equation}
\eta =\tan ^{-1}\frac{y}{x}\,,\quad \sin \alpha =\frac{\sin \eta
}{n}
\end{equation}
and then we can calculate the deflection angle $\delta$
considering the refraction index given by the magnifying glass
model (Eq.(\ref{glass})) or given by the gravitational potential
(\ref{pot}). The mean values, taking into account the Genzel {\it
et al} data \cite{genzel96} are
\begin{equation}
\stackrel{-}{\delta }_{opt}=3.25\pm 0.89\;\mbox{arcsec}\,,\quad
\stackrel{-}{\delta }_{grav}=3.40\pm 1.91\;\mbox{arcsec},
\end{equation}
for the sample of late--type stars, and
\begin{equation}
\stackrel{-}{\delta }_{opt}=3.48\pm 0.58\;\mbox{arcsec}\,,\quad
\stackrel{-}{\delta }_{grav}=3.49\pm 1.71\;\mbox{arcsec},
\end{equation}
for the sample of early-type stars. In both cases the ``optical"
and ``gravitational" results are in a good agreement. In other
words, a magnifying glass model seems to reproduce the spatial
distribution of stars behind Sgr A$^*$. Fig.2 shows such a
spatial distribution for the late and early-type samples.
\begin{figure}[h]
\begin{center}
\epsfig{file=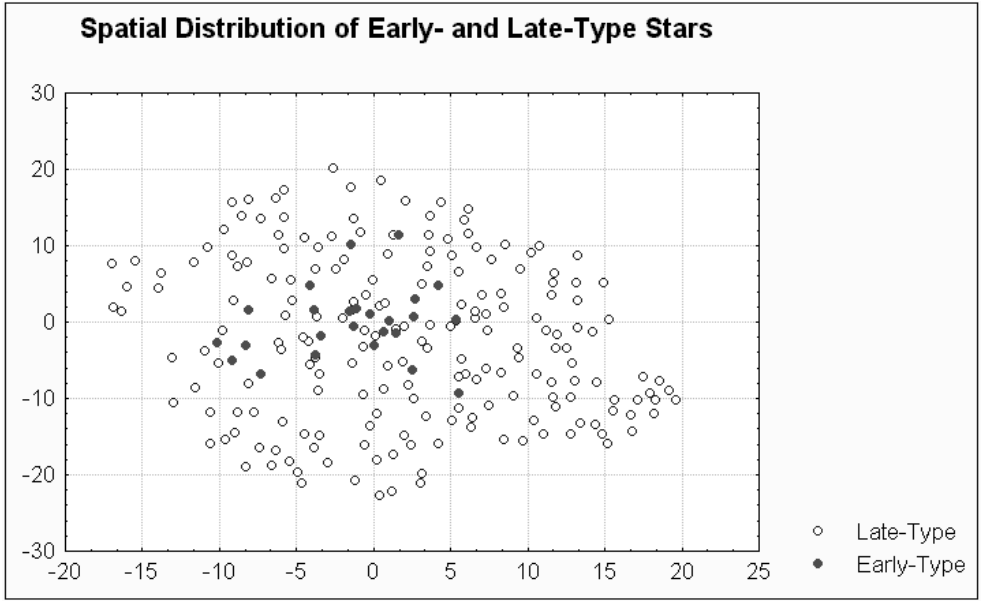,width=13cm,clip=}
\end{center}
\caption{Spatial distribution of the Late-type and Early-type
sample.} \label{fig2}
\end{figure}
The histograms for the deflection angle $\delta $ are given in
Fig.3. It is clear that a correlation exists between the
deflection angle evaluated in both approaches.
\begin{figure}[h]
\begin{center}
\epsfig{file=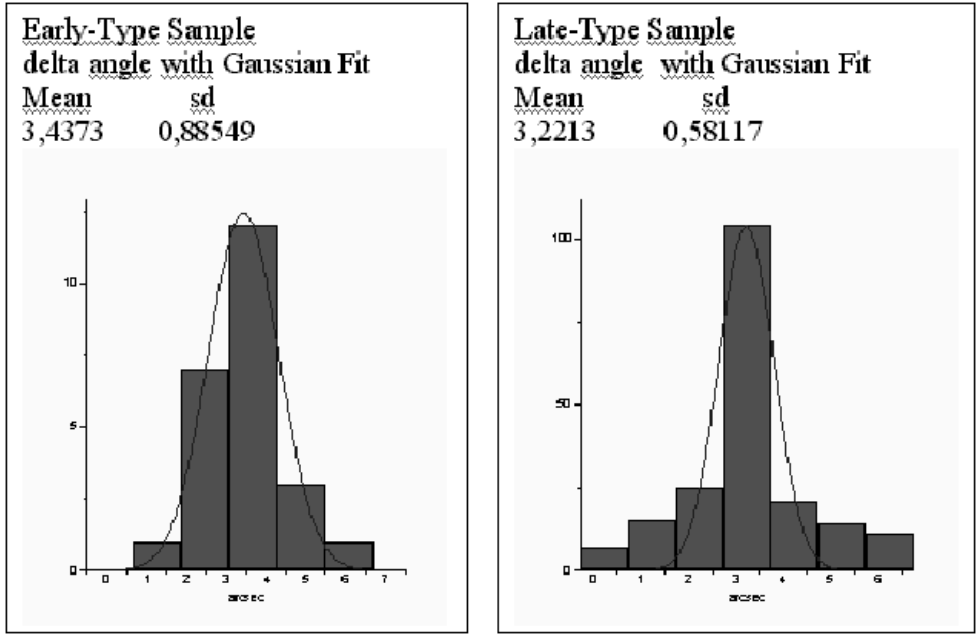,width=13cm,clip=}
\end{center}
\caption{Deflection angle delta for Early- and Late type sample.}
\label{fig3}
\end{figure}
From gravitational lensing point of view, there is no relevant
difference between early and late type samples (see Fig.4).
\begin{figure}[h]
\begin{center}
\epsfig{file=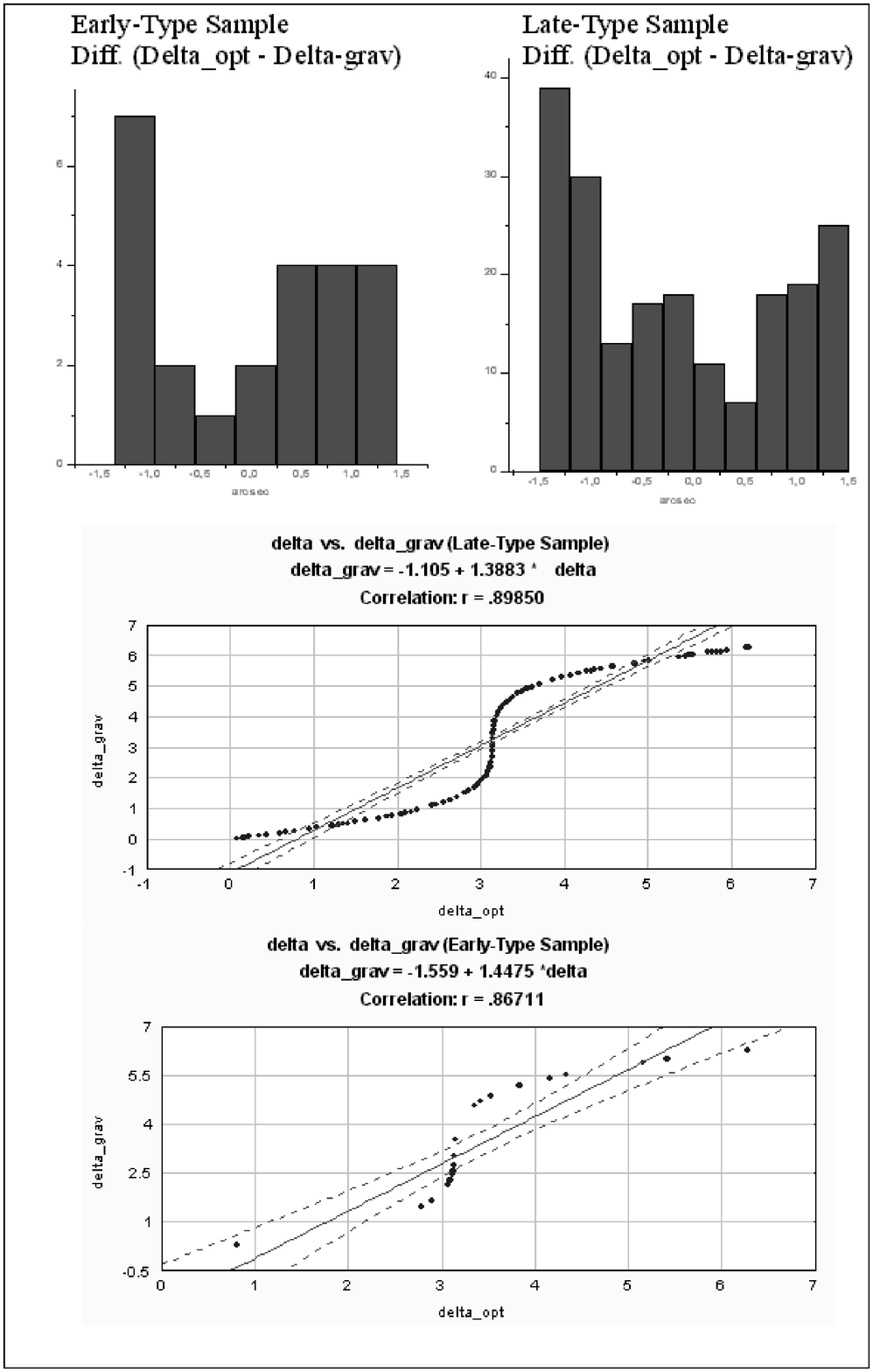,width=13cm,clip=}
\end{center}
\caption{Correlation plots.} \label{fig4}
\end{figure}
A more striking result concerns kinematics. The lensing effect of
a possible neutrino ball at the center of our Galaxy will magnify
stars up to 70$lyrs$ behind, as we showed above, and their
apparent velocities will be larger than in reality, as well (the
effect is similar to that of looking at red fishes moving in a
spherical water--jug of glass!). Taking into account the projected
velocities, they will be corrected by lensing effects, \ie
\begin{equation}
v_{\perp }^{observed}(t)=v_{\perp }^{not\ lensed}(t)+R_{E}/T,
\end{equation}
where
\begin{equation}
R_{E}=\vartheta
_{E}D_{ol}=\sqrt{\frac{4GM}{c^{2}}\frac{D_{ls}}{D_{ol}D_{os}}},
\quad T=\frac{R_{B}}{v_{\perp }^{not\ lensed}}
\end{equation}
are respectively the Einstein radius and the time of crossing.
 $D_{ls}$ is the distance between lens and source; $D_{ol}$ the
distance between observer and lens; $D_{os}$ the distance between
observer and source; $R_{B}$ is the radius of the neutrinos ball
and $M$ its mass. We can assume
\begin{equation}
D_{ls}\approx D_{os}\approx D_{os}/2\approx L_{foc}=f\,.
\end{equation}
We get $R_{E}\propto f^{1/2}$ and then
\begin{equation}
v_{\perp }^{observed}(t)=v_{\perp }^{not\ lensed}(t)\left( 1+\frac{\sqrt{%
\frac{2GMf}{c^{2}}}}{R_{B}}\right) \simeq \left( 1+1.9\right)
\cdot v_{\perp }^{not\ lensed}(t)
\end{equation}
Fig.5 shows $v_{\perp }^{observed}(t)$ as measured by Genzel {\it
et al.} \cite{genzel96}, while the result of our evaluation is  in
Fig.6. It is clear that if we take into account a lensing effect,
the velocity dispersion (\eg the sigma into the plots) of
early--type stars becomes smaller and comparable with the
late-type ones, considering that the standard deviation measured
and reported in \cite{genzel96} is about $30\,Km/s$.
\begin{figure}[h]
\begin{center}
\epsfig{file=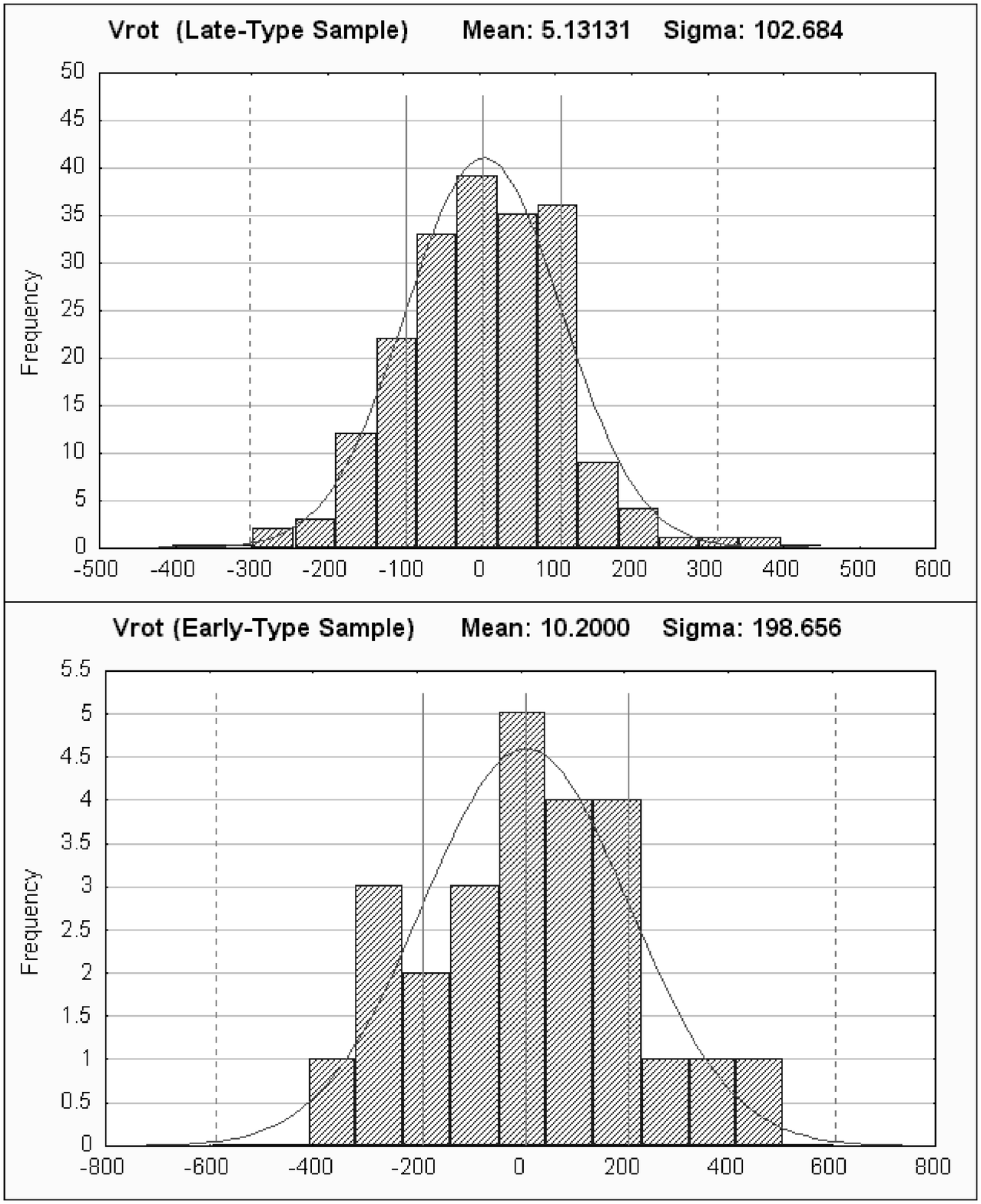,width=13cm,clip=}
\end{center}
\caption{Histograms of observed velocities} \label{fig5}
\end{figure}

\begin{figure}[h]
\begin{center}
\epsfig{file=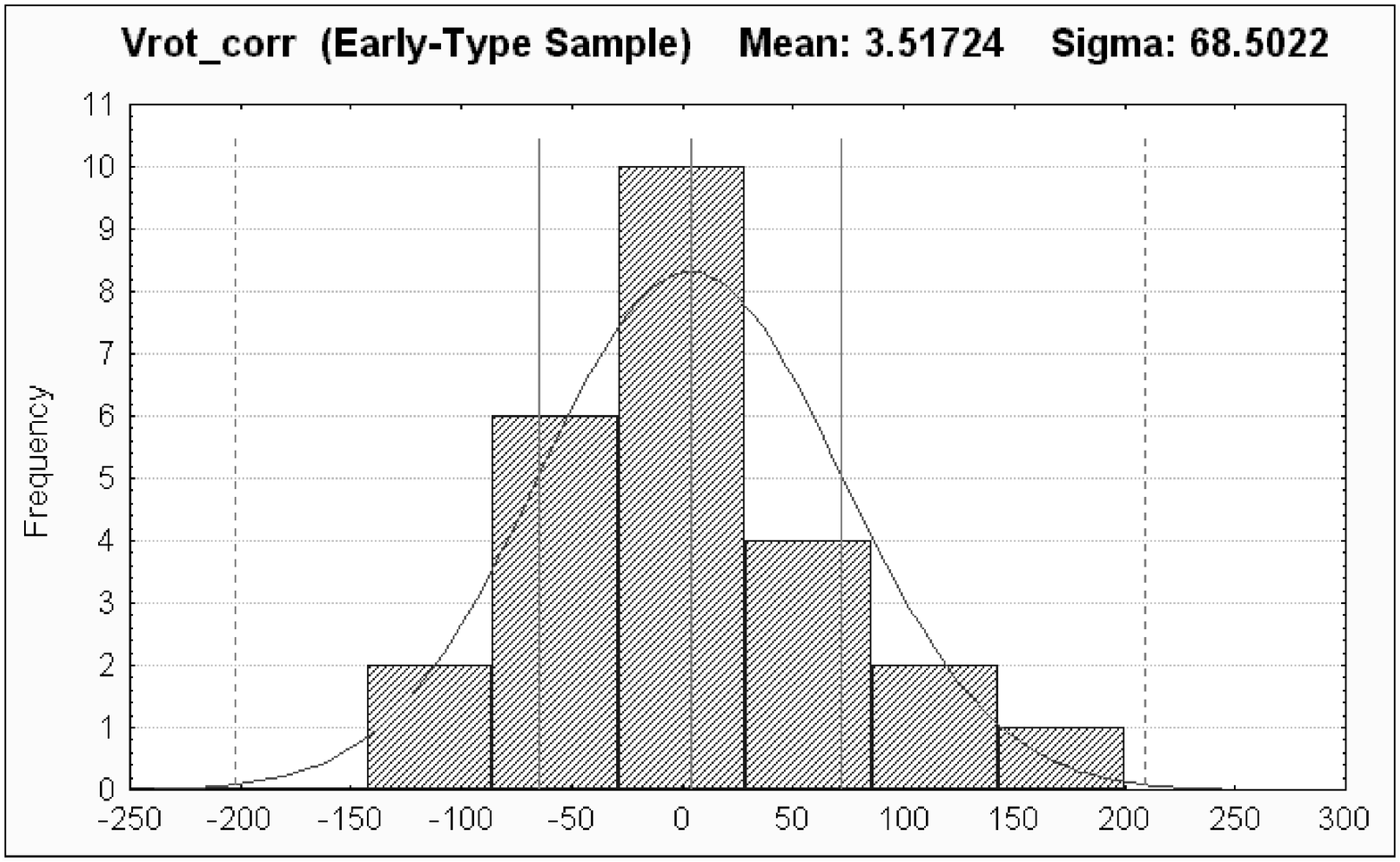,width=13cm,clip=}
\end{center}
\caption{Histograms evaluated velocities.} \label{fig6}
\end{figure}

In conclusion, using such a magnifying neutrino ball, late and
early--type stars could not have different spatial and kinematical
distributions, the only difference should be if they are
``behind'' or ``nearby'' the neutrino ball from the observer point
of view. In fact, as we have shown, early type--stars undergo a
major lensing effect. If these considerations work, the central
compact object could be investigated in such an alternative way
and accurate kinematical and photometric data could give final
answers on its size and nature.

\vspace{ .5 cm}

\noindent {\bf Acknowledgments}\newline The authors are grateful
to G. Scarpetta and R.D. Viollier for fruitful suggestions and
discussions which allow to improve the paper.

\vfill

\end{document}